\documentclass{aastex}          %% The manuscript based on AASTeX v5.x
\usepackage{spr-astr-addons}    %% mimicing ASTR journal style
\begin{document}
%% Article title
%
\title{Contributions to the cross shock electric field at supercritical perpendicular shocks: Impact of the pickup ions}

%% Running heads
\shorttitle{ELECTRIC FIELD AT NONSTATIONARY SHOCKS}
\shortauthors{Yang et al.}

%% Author and Affilations
\author{Zhongwei Yang\altaffilmark{}} \and \author{Desheng Han\altaffilmark{}} \and \author{Huigen Yang\altaffilmark{}} \and \author{Hongqiao Hu\altaffilmark{}} \and \author{Beichen Zhang\altaffilmark{}} \and \author{Qinhe Zhang\altaffilmark{}} \and \author{Ruiyuan Liu\altaffilmark{}}
\affil{SOA Key Laboratory for Polar Science, Polar Research Institute of China, Shanghai, China.}
\email{zwyang@mail.ustc.edu.cn} %% non-output

%% Alternate Affilations
%\altaffiltext{1}{}
%\altaffiltext{2}{}
%\altaffiltext{3}{}

%% Abstract
\begin{abstract}
A particle-in-cell code is used to examine contributions of the pickup ions (PIs) and the solar wind ions (SWs) to the cross shock electric field at the supercritical, perpendicular shocks. The code treats the pickup ions self-consistently as a third component. Herein, two different runs with relative pickup ion density of 25\% and 55\% are presented in this paper. Present preliminary results show that: (1) in the low percentage (25\%) pickup ion case, the shock front is nonstationary. During the evolution of this perpendicular shock, a nonstationary foot resulting from the reflected solar wind ions is formed in front of the old ramp, and its amplitude becomes larger and larger. At last, the nonstationary foot grows up into a new ramp and exceeds the old one. Such a nonstationary process can be formed periodically. When the new ramp begins to be formed in front of the old ramp, the Hall term mainly contributed by the solar wind ions becomes more and more important. The electric field $E_x$ is dominated by the Hall term when the new ramp exceeds the old one. Furthermore, an extended and stationary foot in pickup ion gyro-scale is located upstream of the nonstationary/self-reforming region within the shock front, and is always dominated by the Lorentz term contributed by the pickup ions; (2) in the high percentage (55\%) pickup ion case, the amplitude of the stationary foot is increased as expected. One striking point is that the nonstationary region of the shock front evidenced by the self-reformation disappears. Instead, a stationary extended foot dominated by Lorentz term contributed by the pickup ions, and a stationary ramp dominated by Hall term contributed by the solar wind ions are clearly evidenced. The significance of the cross electric field on ion dynamics is also discussed.
\end{abstract}

%% Keywords

%%  Please use labels (\label, \ref) for section, figure, table,
%%  equation  reference. Use \cite for bibliography references.
%
%\section{}%\label{s:?}
%\subsection{}%\label{ss:?}
%\subsubsection{}%\label{sss:?}

\section{Introduction}

Collisionless shocks are of great interests in space physics, plasma physics and astrophysics since many theoretical, numerical and experimental studies have already evidenced that in the shock transition the bulk flow energy of plasmas is converted irreversibly into thermal energy \citep{sagdeev1966,lembege2004,burgess2005}. During the conversion process, the interaction between the electromagnetic fields and particles replaces the role played by collisions in a normal hydrodynamics. In quasi-perpendicular shocks ($\theta_{Bn}>45^\circ$, $\theta_{Bn}$ is the angle between the upstream magnetic field and shock normal), the normal/cross shock electric field/potential plays an key role in both acceleration \citep{zank1996,shapiro2003,yang2009a} and thermalization \citep{burgess1989,lembege1992} of the incident particles. In-situ observations of the electric field at Earth's bow shock provide important and unique experimental data that can be used to understand the details of plasma thermalization and energetic particle production in more general circumstances. \cite{walker2004} experimentally studied the structure of the electric field profile during a number of crossings of the quasi-perpendicular bow shock using data from the Cluster satellites. Their results showed that the amplitude of the cross shock electric field appears to increase as $\theta_{Bn}$ approaching $90^\circ$. Based on Cluster four-spacecraft observations, contributions to the electric field of the terrestrial bow shock are analyzed in detail by \cite{eastwood2007}. They found that the Hall term in the generalized Ohm's law accounts for a majority of the normal electric field of supercritical, quasi-perpendicular shocks by two different methods: 1. the single-spacecraft technique in which the shock is assumed as one-dimensional (1-D) and stationary, and 2. the curlometer technique \citep{dunlop2002} which requires that the spacecraft are co-located in the shock ramp and the shock is stationary among the satellites.

Both particle-in-cell(PIC) and hybrid simulations have clearly evidenced that supercritical, quasiperpendicular shocks are nonstationary and suffer a self-reformation on the local gyro scale of the incident ions \citep{lembege1987,Nishimura2003,Scholer2004,Hellinger2007,lembege2009,yang2011a,Shinohara2011}. \cite{hada2003} analyzed the time evolution of the normal electric field and the fraction of reflected ions at the reforming shock front by using both PIC simulation and semi-numerical model. Their findings suggest that the reflected ions have a very coherent motion (narrow ring) as described by the so called trapping loop/vortex in the phase space ($V_{x}-X$) at the beginning of the reflection. This coherent forces these ions to accumulate over a narrow spatial range at a foot length distance from the ramp; it results that ramp and foot are well separated. This accumulation is suitable for building up a peaked-like normal electric field which itself reflects at later times new incoming ions at the foot before these reach the ramp. The peak of electric field produces an enhancement of local magnetic field. At last, a new ramp (i.e. grown-up foot) is fully built up and substitute the old one. This process is periodic, and is named self-reformation. Nonstationary characteristic of the shock front is later confirmed by an individual case study of Cluster observation at the terrestrial bow shock \citep{lobzin2007}. \cite{Mazelle2010} statistically confirms this robust characteristic. In addition, they found the magnetic field ramp of the shock often reaches a few $c/\omega_{pe}$, and the time variability of scales observed for the same upstream conditions is also a clear signature of the nonstationary shock front.

In recent study on collisionless magnetic reconnection. First, at the ion inertial length scale, ions and electrons are decoupled \citep{Pritchett2001,Lu2011}. The resulting Hall effect determines the reconnection electric field. Second, at the electron inertial scale, even the electrons are demagnetized, and the electric field is dominated by the off-diagonal electron pressure \citep{Vasyliunas1976,Pritchett2001}. In order to understand the physics in different scales (e.g. the whole shock front is in ion inertial scale and the ramp is in electron inertial scale) of the shock front structures of a supercritical, perpendicular shock, \cite{yang2009b} borrowed the same techniques which has been successfully applied to the research on collisionless magnetic reconnection. First, a PIC simulation including all of the behaviors of ions and electrons were performed. Nonstationary fields and corresponding particle dynamics were obtained self-consistently. Second, contributions of Lorentz, Hall and electron pressure terms to the cross shock electric field were calculated by using the self-consistent particle data issued from the PIC simulation. The main conclusions are the following: (1) the contributions of Lorentz, Hall and electron pressure terms to the cross shock electric field is evolving versus time. The electron pressure term within the shock front always contributes little; (2) when the new ramp begins to be formed in front of the old ramp, the Hall term becomes more and more important; (3) the electric field $E_x$ is dominated by the Hall term when the new ramp exceeds the old one. Nevertheless, the shock front topology of a nonstationary shock can be modified by the local pickup ions (PIs).

Pickup ions are extensively and directly observed in the solar wind upstream of the Earth's bow shock (at $\sim$1 AU) with the SULEICA instrument on aboard AMPTE/IRM \citep{mobius1985,Oka2002}, around the shock associated with the corotating interaction region (CIR) in the interplanetary space (at $\sim$4.5AU) with SWICS instrument onboard the Ulysses spacecraft \citep{Gloeckler1994}, and in the upstream of the Halley comet bow shock with HERS instrument on the Giotto spacecraft \citep{Neugebauer1989}. The production process and feature of the pickup ions in the solar wind have been studied quite well \citep{Sagdeev1986,Lee1987}. Neutral atoms and molecules, penetrating from the local interstellar medium or escaping from comets, are ionized by photoionization, electron impact or charge exchange with the solar wind. Under the influence of the solar wind magnetic field the ions compose a ring-beam distribution in velocity space. Then they are scattered by ambient and excited Alfv\'{e}nic fluctuations to form a spherical shell distribution. Interstellar neutral hydrogen ($\rm{H^+}$) has a much higher density than other interstellar neutrals; interstellar helium accounts for $\sim10\%$ of the interstellar gas and its ionization rate is $\sim10\%$ that of hydrogen \citep{mobius1986}. The percentage of pickup ions to the solar wind ions (SWs) at 50 AU is estimated to be $10\%$ and increase linearly with distance from the sun \citep{Vasyliunas1976}. So the pickup $\rm{H^+}$ is expected to be the dominant (i.e. number density of pickup protons is $>20\%$ that of the solar wind protons) pickup species at the termination shock (TS) at about $70\sim100$ AU \citep{Richardson2008,Wu2009}. Early 1-D hybrid simulations \citep{Omidi1986,Omidi1987,Liewer1993} have revealed that an extended foot, which scales in reflection ion gyroradius, is seen on the supercritical shocks as a results of the reflected pickup ions. \cite{Matsukiyo2011} examined the effect of pickup ions on the heliospheric termination shock by using a 1-D PIC code. They found the pickup ions in the extended foot causes an electric field in the shock normal direction. This leads to a large increase of the shock potential barrier upstream of the magnetic field ramp. However, the impact of pickup ions on the self-reformation process and contributions to the cross shock electric field at nonstationary shocks has not been analyzed yet. This gave us the motivation for investigating the impact of PIs on the relative contributions of PIs and solar wind ions (SWs) respectively to the total normal electric field of the shock front.

This paper represents an extension of a previous work \citep{yang2009b}. Herein, we aim to address the following questions by using one-dimensional PIC simulation (the code treats the pickup ions self-consistently as a third component): (1) Which term (such as Hall term and Lorentz term) in the generalized Ohm's law accounts for a majority of the normal electric field of a supercritical, perpendicular shock in the presence of pickup ions? Does it vary versus time? (2) What is the contributions of pickup ions and solar wind ions to the terms mentioned above? (3) How is this impact according to different percentages of upstream pickup ions on this competition? (4) What are the consequences of high percentage of pickup ions on the nonstationarity of the shock front? This paper is organized as follows. In section 2, we briefly describe the numerical model used herein. Sections 3 and 4 present the simulation results at (a) low PIs percentage case, and (b) high PIs percentage case, respectively. The main conclusions will be summarized in section 5.

\section{Simulation Conditions}

We use a one-dimensional electromagnetic PIC code to simulate the evolving structure of a supercritical, collisionless, perpendicular shock. Simulations of nonstationary shocks have been performed, for example, by \cite{yang2011b}, where the pickup ions ($\rm{H^+}$) are treated as test particles. Herein, the shock is produced by the injection method as in previous works \citep{Quest1985,burgess1989,Nishimura2003,Chapman2005}. In order to self-consistently reproduce the impacts of pickup ions, we simulate three species within our code: electrons, solar wind ions, and pickup ions. Particles are injected on the left-hand side of the simulation box with a inflow/upstream drift speed $V_{inj}$, and are reflected at the other end. The distribution functions for the solar wind ions and electrons are Maxwellian. Pickup ions are injected on a thin sphere in velocity space centered at $V_{inj}$ with radius $V_{shell}$ as in earlier works \citep{Kucharek1995,Lee2005,Wu2009,Matsukiyo2011}. The shock built up by an ion/ion beam instability \citep{Scholer2004} and moves with a speed $V_{ref}$ from the right-hand side toward the left. The upstream Alf\'{e}nic Mach number of the shock is $M_A=(V_{inj}+V_{ref})/V_{A}$ , where the Alf\'{e}n speed $V_A$ is equal to 1. All basic parameters are as follows: Plasma box size length $L_{x}=92.16(c/\omega_{pi})$; velocity of light $c=21.4$, and mass ratio $m_{i}/m_{e}=1836$. The electron/ion temperature ratio $T_e/T_i=1.36$ is chosen in order to investigate the particle acceleration at a reforming shock. The ambient magnetic field along the $y$ direction is $|B_{0}|=1$. Low beta values $\beta_i=0.125$ and $\beta_e=0.1$ are chosen in order to investigate the particle acceleration at a reforming shock. Initially there are 200 particles of each species in a cell. The upstream plasma is quasi-neutrality, i.e. $n_e=n_i$, where $n_i=n_{SWs}+n_{PIs}=N_{SWs}\times {\rm SWs}\%+N_{PIs}\times {\rm PIs}\%$. $n_e$, $n_i$, $n_{SWs}$ and $n_{PIs}$ are densities of the electrons, the total ions, the solar wind ions and the pickup ions, respectively. $N_{SWs}$, $N_{PIs}$, $SWs\%$ and $PIs\%$ are the counts of each species of ions and their relatively weighted percentage in the PIC simulation. For these initial conditions, the plasma parameters are summarized in Table \ref{table:1} for both electrons and SWs. The Larmor gyro radius in the table is calculated based on the thermal velocity.

\begin{table}[t]
\caption{Upstream plasma parameters defined for the 1-D PIC simulation}
\label{table:1}
\begin{tabular}{@{}lcccccccc@{}}
\tableline
& & &  Electrons & SWs \\
\tableline
 Thermal velocity    &     $V_{th x,y,z}$                & & 15            & 0.3   \\
 Larmor gyro radius  &     $\rho_{c}$                    & & 0.0082        & 0.3   \\
 Inertia length      &     $c/\omega_{p}$                & & 0.0233        & 1     \\
 Gyro frequency      &     $\Omega_{c}$                  & & 1836          & 1     \\
 Plasma frequency    &     $\omega_{p}$                  & & 918           & 21.4  \\
 Gyro period         &     $\tau_{c}$                    & & 0.0034        & 6.28  \\
 Plasma beta         &     $\beta$                       & & 0.1225        & 0.1   \\
\tableline
\end{tabular}
\end{table}

\section{Simulation Results}

This section can be separated into two parts. In the first part, we show the contributions of solar wind ions and pickup ions to the cross shock electric fields at three typical shock profile issued from the low PIs\% case. In this case, the percentage of the pickup ions is 25\%, and the shock front is characterized by a self-reformation process with a cyclic period about 1.36$\Omega_{ci}^{-1}$ ($\Omega_{ci}$ is the upstream ion gyro-frequency). During this reformation, the instantaneous Mach number of the shock at different times can differ from the average value $M_{A}$=5.29. In the second part, we do similar things for high PIs\% case, in which the percentage of the pickup ions is set to 55\%. In this case, the shock front nonstationarity is fully suppressed. In each part, the analysis is based on systematic comparison between the Hall term and Lorentz term (contributed by solar wind ions and/or pickup ions) in the generalized Ohm's law in order to clarify the answers to the questions addressed in this paper and mentioned in section 1.

\subsection{Effect of Low Percentage Pickup Ions on the Cross Shock Electric Field (nonstationary case)}

\begin{figure}%[tb]
\includegraphics[width=\columnwidth]{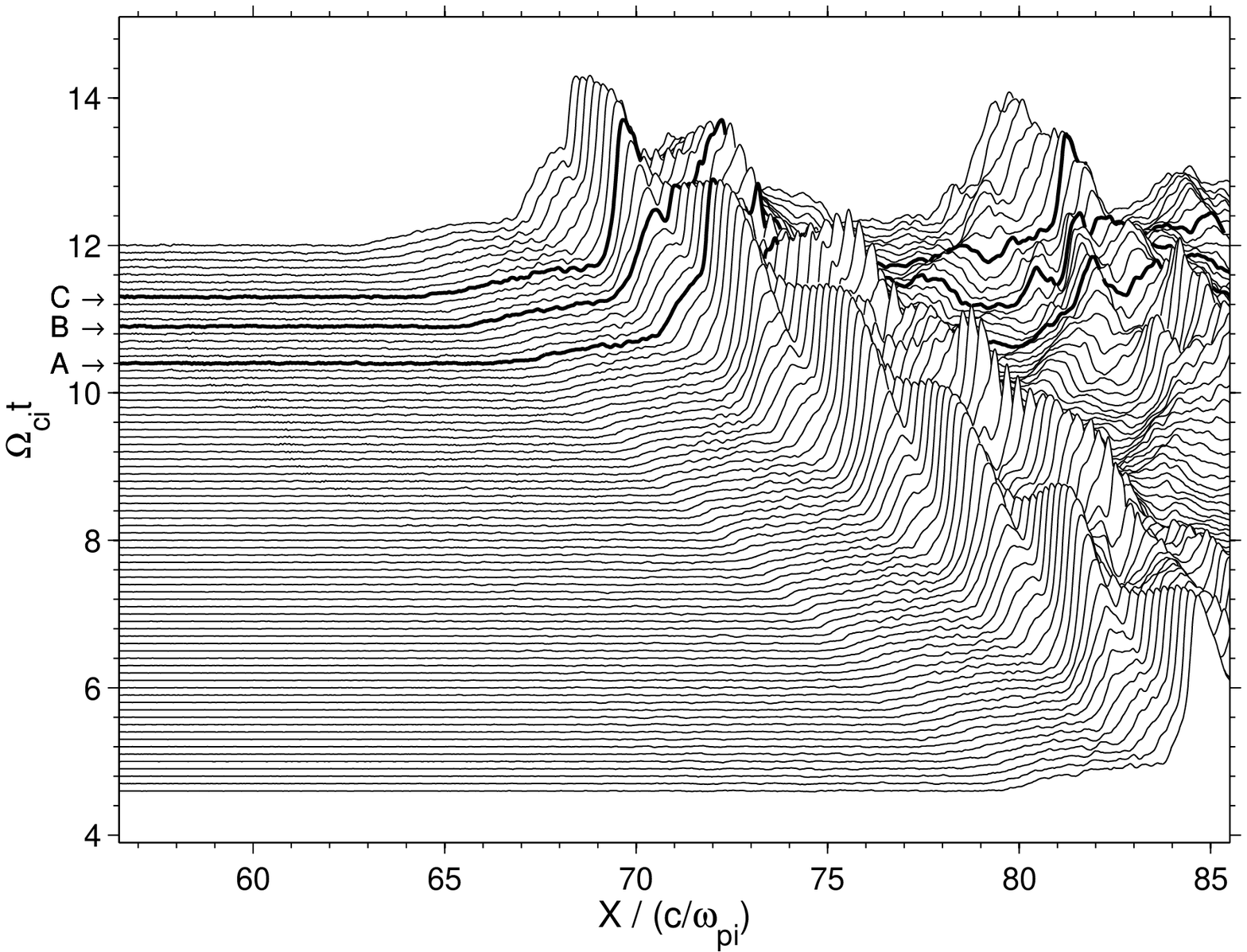}
\caption{The time evolution of $B_y$ versus $X$. Thick curves A, B, and C indicate the shock profiles that are chosen in the stack plots within one cyclic self-reformation of the shock front (with 25\% pickup ions).}\label{fig:1} %% no full stop at the end of caption
\end{figure}

Firstly, let us analyze the impact of low percentage pickup ions on the cross shock electric filed at a nonstationary shock. The nonstationarity of the perpendicular shock can be demonstrated in figure 1, which shows the time evolution of the magnetic field $\tilde{B}_{y}$ from $t=$4.6 to 12$\Omega_{ci}^{-1}$. The shock is propagating from the right to the left. At about $t=$4.6, the shock ramp (``old" ramp) is at about $X=$84.48$c/\Omega_{pi}$. Later at about $t=$5.2$\Omega_{ci}^{-1}$, the foot amplitude becomes larger and larger due to the accumulation of the reflected solar wind ions in the foot. Then a ``new" shock ramp is formed. At about $t=$5.6$\Omega_{ci}^{-1}$, it exceeds the amplitude of the ``old" ramp, and the ``new" shock ramp is around $X=$81.89$c/\Omega_{pi}$. Simultaneously, the ``old" shock ramp becomes weaker and weaker, and is located downstream of the ``new" ramp. The shock front is characterized by a self-reformation with a cyclic period about 1.36$\Omega_{ci}^{-1}$. The self-reformation of the shock front is due to the coupling between the ``incoming" and ``reflected" ions, which has been demonstrated by \cite{hada2003} at supercritical perpendicular shocks. We see that the solar wind ion dynamics proceed essentially as in the absence of the pickup ions, whereas the pickup ions form an stationary extended foot starting $\sim4(c/\Omega_{ci})$ upstream. This phenomenon has already been demonstrated by a good deal of hybrid and PIC simulations \citep{Omidi1986,Liewer1993,Lipatov1999,Chapman2005,Lee2005,Matsukiyo2007,Wu2009,Matsukiyo2011}. In this subsection, we will clarify why the low percentage pickup ions can not suppress the self-reformation of the shock front by analyzing the cross shock electric field which play a key role in the reflection of incident ions.

Let's concentrated on the evolution of the electric field in this nonstationary perpendicular shock. Assuming a two-fluid model for the plasma, the electric field can be obtained directly from the electron momentum equation without further approximation
\begin{eqnarray}\label{eqn:1}
\bold{E}=-\bold{U}_{e}\times\bold{B}-
\frac{\nabla\cdot\underline{\bold{P}}_{e}}{en_{e}}-
\frac{m_{e}}{e}\frac{d\bold{U}_{e}}{dt}.
\\ \nonumber
\end{eqnarray}
Here $\bold{U}_{e}$ is the electron fluid velocity, $\bold{B}$ the magnetic field, $\underline{\bold{P}}_{e}$ the electron pressure tensor, $m_{e}$ the electron mass, and $e$ the electric charge. If we use the total current density $\bold{J}=e(n_{i}\bold{U}_{i}-n_{e}\bold{U}_{e})$ (where $\bold{U}_{i}$ is the ion fluid velocity) and considering the one-dimensional property of our simulation, the normalized Eq. (\ref{eqn:1}) can be rewritten as
\begin{eqnarray}\label{eqn:2}
\nonumber
\bold{E}=-\frac{n_{i}}{n_{e}}\bold{U}_{i}\times\bold{B}+
\frac{1}{n_{e}}\bold{J}\times\bold{B}-
\frac{1}{n_{e}}(\frac{\partial n_{e}T_{exx}}{\partial x}\bold{\hat{e}}_{x}+\\
                         \frac{\partial n_{e}T_{exy}}{\partial x}\bold{\hat{e}}_{y}+
                         \frac{\partial n_{e}T_{exz}}{\partial x}\bold{\hat{e}}_{z})
-\frac{d\bold{U}_{e}}{dt},
\end{eqnarray}
where $T_{exx}=\sum\limits_{k=1}^{n_{e}}(V_{ex,k}-U_{ex})^2/n_e$, $T_{exy}=\sum\limits_{k=1}^{n_{e}}(V_{ex,k}-U_{ex})\times(V_{ey,k}-U_{ey})/n_e$, and $T_{exz}=\sum\limits_{k=1}^{n_{e}}(V_{ex,k}-U_{ex})\times(V_{ez,k}-U_{ez})/n_e$ are components of the electron kinetic temperature \citep{Heinz2005}. $\bold{\hat{e}}_{x}$, $\bold{\hat{e}}_{y}$, and $\bold{\hat{e}}_{z}$ are unit vectors. The first, second, third and fourth terms on the right of Eq. (\ref{eqn:2}) are Lorentz term, Hall term, electron pressure gradients term and electron inertial term, respectively. All of the physical quantities on the right of Eq. (\ref{eqn:2}) can be statistically obtained from the PIC simulation. We choose three typical shock profiles at three different times (marked by ``A", ``B" and ``C" in figure 1) within one shock self-reformation cycle (from $t=$10.1$\Omega_{ci}^{-1}$ to 11.4$\Omega_{ci}^{-1}$) and analyze separately the corresponding roles of Lorentz and Hall terms in the cross shock electric field $E_x$. Herein, we concentrate on the competition of the solar wind ions and pickup ions for Lorentz and Hall terms during the shock front self-reformation. The electron pressure gradients term and electron inertial term are neglected because they do not play important roles in ion reflection and shock nonstationarity \citep{yang2009b}. Since the normalized bulk velocities of the total ions, SWs and PIs are $\bold{U}_i=(\sum\limits_{k=1}^{N_{SWs}}\bold{V}_{SWs,k}\times{\rm SWs}\%+\sum\limits_{k=1}^{N_{PIs}}\bold{V}_{PIs,k}\times{\rm PIs}\%)/n_i$, $\bold{U}_{SWs}=\sum\limits_{k=1}^{N_{SWs}}\bold{V}_{SWs,k}/n_{SWs}$ and $\bold{U}_{PIs}=\sum\limits_{k=1}^{N_{PIs}}\bold{V}_{PIs,k}/n_{PIs}$, respectively. The normalized current densities contributed by the total ions, SWs and PIs can be written as $\bold{J}=n_i\bold{U}_i-n_e\bold{U}_e$, $\bold{J}_{SWs}=n_{SWs}\bold{U}_{SWs}-\dfrac{n_{SWs}}{n_i}n_e\bold{U}_e$ and $\bold{J}_{PIs}=n_{PIs}\bold{U}_{PIs}-\dfrac{n_{PIs}}{n_i}n_e\bold{U}_e$, respectively. Eq. (\ref{eqn:2}) becomes:
\begin{eqnarray}\label{eqn:3}
\nonumber
E_x=-\frac{n_{SWs}}{n_{e}}(\bold{U}_{SWs}\times\bold{B})_x
-\frac{n_{PIs}}{n_{e}}(\bold{U}_{PIs}\times\bold{B})_x+\\
\frac{1}{n_{e}}(\bold{J}_{SWs}\times\bold{B})_x+
\frac{1}{n_{e}}(\bold{J}_{PIs}\times\bold{B})_x,
\end{eqnarray}
where the first and second terms on the right of Eq. (\ref{eqn:3}) are contributions of SWs and PIs to the Lorentz term of $E_x$, and the other two are contributions of SWs and PIs to the Hall term of $E_x$.

\begin{figure}%[tb]
\includegraphics[width=\columnwidth]{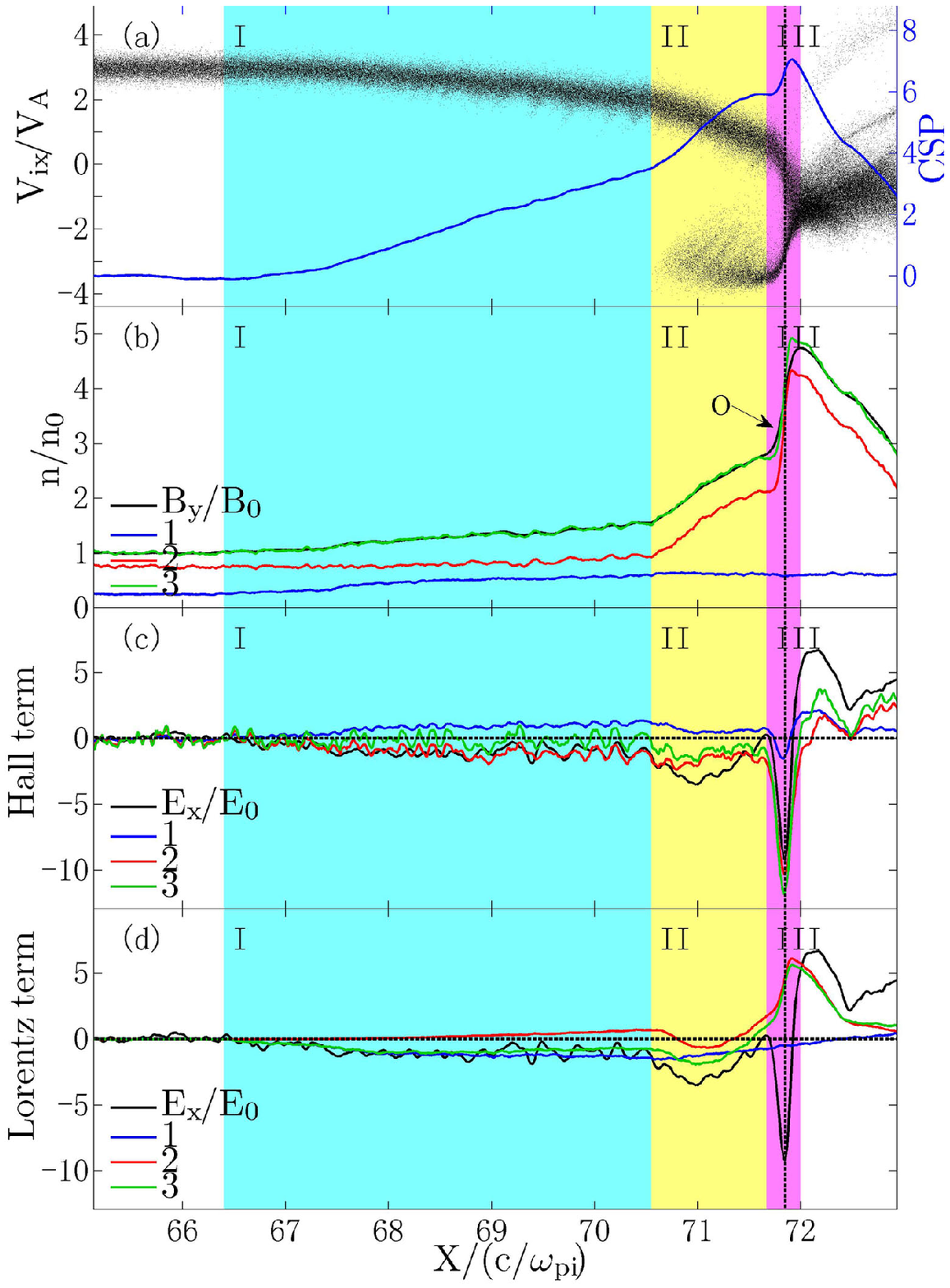}
\caption{(a) phase space plots ($V_{ix}-X$) of the solar wind ions at the shock profile A ($t=$10.4$\Omega_{ci}^{-1}$). And the cross shock potential is also shown for reference (in blue). (b) the density profiles of the pickup ions, solar wind ions and their total are in blue (line 1), red (line 2) and green (line 3), respectively. The main magnetic field $B_y$ (black curve) is also shown for reference. ``O" denotes the position of the old ramp (marked by a vertical dashed) during one reforming cycle. (c) the cross shock electric field $E_x$ is indicated by the black curve. The additional curves indicate the contributions to $E_x$ of the Hall terms on the right of Eq. (\ref{eqn:3}): pickup ion Hall term $\frac{1}{n_{e}}(\bold{J}_{PIs}\times\bold{B})_x$ (line 1), solar wind ion Hall term $\frac{1}{n_{e}}(\bold{J}_{SWs}\times\bold{B})_x$ (line 2), and their total (line 3). (d) corresponding plots for the Lorentz terms on the right of Eq. (\ref{eqn:3}): pickup ion Lorentz term $-\frac{n_{PIs}}{n_{e}}(\bold{U}_{PIs}\times\bold{B})_x$ (line 1), solar wind ion Lorentz term $-\frac{n_{SWs}}{n_{e}}(\bold{U}_{SWs}\times\bold{B})_x$ (line 2), and their total (line 3). The cyan, yellow and magenta highlight the regions: stationary foot, nonstationary foot and the nonstationary ramp, respectively.}\label{fig:2} %% no full stop at the end of caption
\end{figure}

Figure 2a-2c shows the snapshots of CSP (cross shock potential), $B_y$ and $E_x$ (in black) at shock profile A (at $t=10.4\Omega_{ci}^{-1}$) of a reforming shock, respectively. At this time, the shock front can be divided into three regions: region I (highlighted in cyan) is the stationary foot dominated by the pickup ions, region II (highlighted in yellow) indicates the time-varying/nonstationary foot results from the accumulation of the reflected solar wind ions, region III (highlighted in magenta) is the nonstationary shock ramp. The position of the ramp ``O" (denoted by vertical dashed) is at $X=$71.85$c/\omega_{pi}$. The width of the shock ramp is measured from the maximum point of the cross shock electric field $-E_x$ to the maximum point of the magnetic overshoot \citep{Shimada2005,yang2011a}, is about 0.15$c/\omega_{pi}=$6.4$c/\omega_{pe}$. And the widths of foots in region I and II are about 4$c/\omega_{pi}$ and 1.3$c/\omega_{pi}$, respectively. Figure 2a shows the phase space plots ($V_{ix}-X_i$) of the solar wind ions. A fraction of the them are reflected at region III and accumulated at region II. Corresponding phase space plot of the pickup ions is not shown here, and it has already been analyzed in detail by \cite{Matsukiyo2011}. On this point, there is no longer to do more. Figure 2b shows the density profiles of the PIs (blue), the SWs (red) and the total ions (green). In region I, the contribution of PIs is about two-thirds of that of the SWs. The density profile of SWs in this region is flat and the density of PIs increases slightly. In region II, the density of SWs increases distinctly, and the density of PIs becomes flat. And in region III, the local density profile of SWs increases precipitously because most of the incident solar wind ions are reflected at the ramp, and the density profile of the PIs is still flat. In figure 2c, the contributions of the PIs, SWs and the total ions to the Hall term of the cross shock electric field $E_x$ (black curve) are indicated by blue, red and green curves, respectively. The main features are summarized as follows: 1. in region I, the amplitude of the total Hall term is close to zero. Because the contributions of PIs and SWs to the total Hall term (in green) are roughly equal and opposite to each other. 2. in region II, the Hall effect of the PIs (i.e. the term $\frac{1}{n_{e}}(\bold{J}_{PIs}\times\bold{B})_x$ in Eq. (\ref{eqn:3}), in blue) is almost zero, and the Hall term is dominated by SWs. 3. in region III, the Hall effect associated with the SWs is much larger than that of the PIs. The spatial form of the Hall term contributed by SWs (i.e. the term $\frac{1}{n_{e}}(\bold{J}_{SWs}\times\bold{B})_x$ in Eq. (\ref{eqn:3}), in red) apparently matches with that of $E_x$. In figure 2d, the contributions of the PIs, the SWs and the total ions to the Lorentz term of the cross shock electric field $E_x$ (black curve) are color-coded severally as in figure 2c. The main features are as follows: 1. in region I, the effect of SWs on the total Lorentz term (in green) is very small. The PIs dominated the Lorentz term in this region. 2. in region II, the contributions of SWs (i.e. the term $-\frac{n_{SWs}}{n_{e}}(\bold{U}_{SWs}\times\bold{B})_x$ in Eq. (\ref{eqn:3}), in red) is a bit smaller than that of PIs (i.e. the term $-\frac{n_{PIs}}{n_{e}}(\bold{U}_{PIs}\times\bold{B})_x$ in Eq. (\ref{eqn:3}), in blue) to the total Lorentz term. 3. in region III (i.e. in the vicinity of the shock ramp), the contribution of PIs to the Lorentz term is almost zero.

In summary, the cross shock electric field $E_x$ in region I (i.e. at the stationary extended foot) is dominated by the Lorentz term, and is mainly contributed by PIs. In region II (i.e. at the time-varying solar wind ion foot, which will be clarified in the remainder of this subsection), the contributions of the Lorentz term and Hall term to the total $E_x$ are comparable to each other. And the profile $E_x$ in region III (i.e. at the steep ramp) is dominated by the Hall term, and is mainly contributed by SWs.

\begin{figure}%[tb]
\includegraphics[width=0.95\hsize]{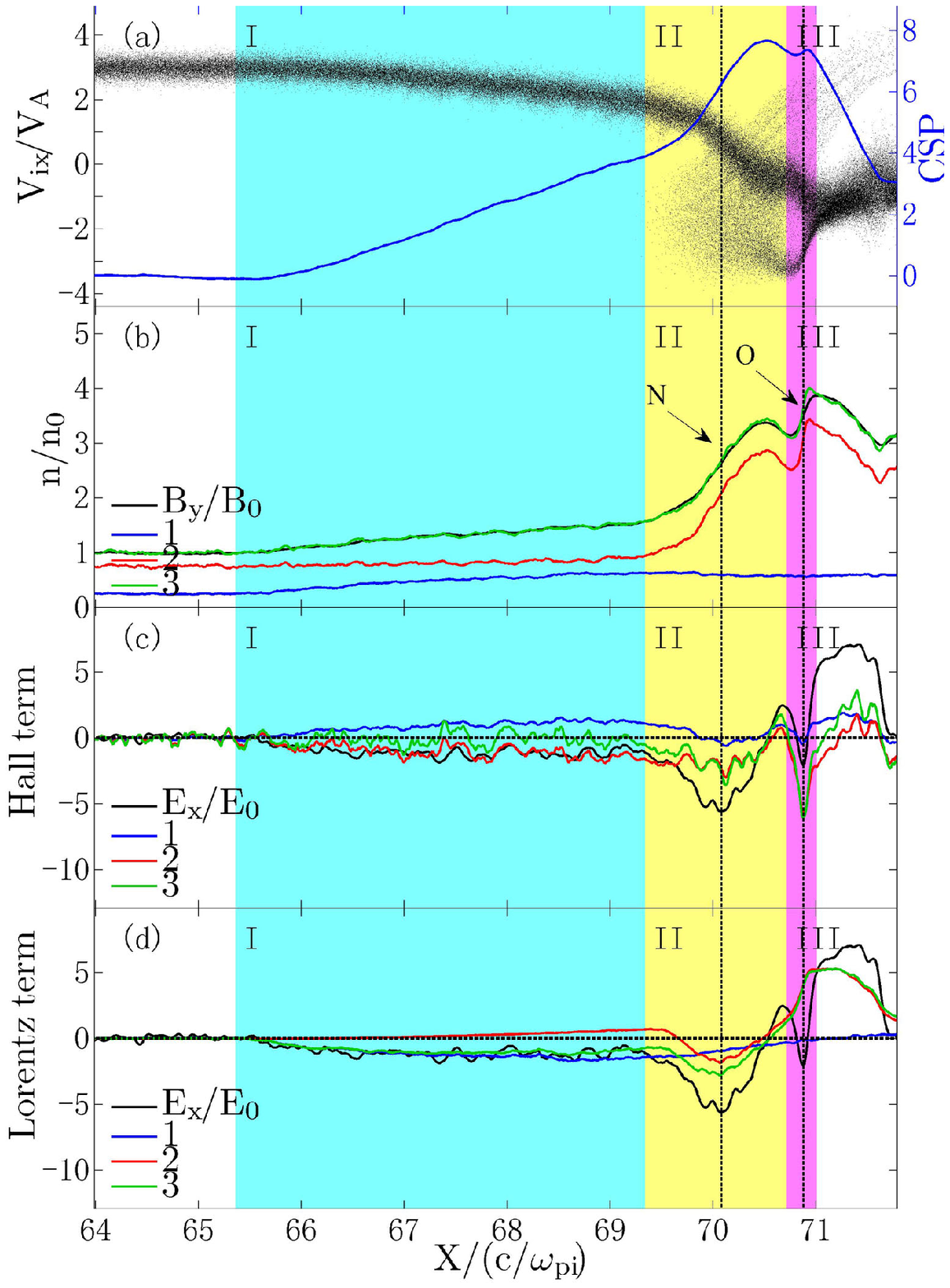}
\caption{(a) phase space plots ($V_{ix}-X$) of the solar wind ions at the shock profile A ($t=$10.9$\Omega_{ci}^{-1}$). And the cross shock potential is also shown for reference (in blue). (b) the density profiles of the pickup ions, solar wind ions and their total are in blue (line 1), red (line 2) and green (line 3), respectively. The main magnetic field $B_y$ (black curve) is also shown for reference. ``O" and ``N" denote the positions of the old and new ramps (marked by a vertical dashed lines) during one reforming cycle. (c) the cross shock electric field $E_x$ is indicated by the black curve. The additional curves indicate the contributions to $E_x$ of the Hall terms on the right of Eq. (\ref{eqn:3}): pickup ion Hall term $\frac{1}{n_{e}}(\bold{J}_{PIs}\times\bold{B})_x$ (line 1), solar wind ion Hall term $\frac{1}{n_{e}}(\bold{J}_{SWs}\times\bold{B})_x$ (line 2), and their total (line 3). (d) corresponding plots for the Lorentz terms on the right of Eq. (\ref{eqn:3}): pickup ion Lorentz term $-\frac{n_{PIs}}{n_{e}}(\bold{U}_{PIs}\times\bold{B})_x$ (line 1), solar wind ion Lorentz term $-\frac{n_{SWs}}{n_{e}}(\bold{U}_{SWs}\times\bold{B})_x$ (line 2), and their total (line 3). The cyan, yellow and magenta highlight the regions: stationary foot, nonstationary foot and the nonstationary ramp, respectively.}\label{fig:3} %% no full stop at the end of caption
\end{figure}

Figure 3a-3c shows the snapshots of CSP (cross shock potential), $B_y$ and $E_x$ (in black) at shock profile B (at $t=10.9\Omega_{ci}^{-1}$) of a reforming shock, respectively. At this time, the whole shock front still can be divided into three regions as in figure 2. The amplitude of the old ramp ``O" is decreased, and its width is about 0.13$c/\omega_{pi}=$5.6$c/\omega_{pe}$. It located at $X=$70.88$c/\omega_{pi}$. The extended foot caused by pickup ions in region I is stationary, and the nonstationary foot caused by solar wind ions in region II becomes broader. Their widths are about 4$c/\omega_{pi}$ and 1.55$c/\omega_{pi}$, respectively. Now, the cross electric field $\tilde{E}_{x}$ still has a negative value at the shock old ramp (``O"). However, its amplitude is strongly decreased and the foot in region II grows into a new ramp (``N"). Figure 3a shows the phase space plots ($V_{ix}-X_i$) of the solar wind ions. It clearly appears that the reflected ions have a very coherent motion (within regions II and III) as described by the well defined trapping loop (vortex). This vortex is a manifestation of shock front self-reformation [Hada et al., 2003]. At this time, the bulk velocity of the incident solar wind ions begin to decrease at the new (``N") ramp in region II instead the old one in region III. Figure 3b shows the density profiles of the PIs (blue), SWs (red) and the total ions (green), as in figure 2b. In region I, the total density profile gradually increases versus $X$ due to the accumulation of the reflected pickup ions. More and more solar wind ions, which are reflected at the ramp in early time, are accumulated in region II. Thus a new ramp (``N") appears and its amplitude reaches at about 87.5\% of the old ramp (``O"). The reflected solar wind ions feeds up the self-reformation of the shock front. And in region III, the density profile of the SWs is decreased. The local electromagnetic fields are weaker than that at profile A. The density profile of SWs in region II and III have a relationship of restricting each other. And the density profiles of PIs in region II and III are still flat as that in figure 2b. The main features shown in figure 3c are summarized as follows: 1. in region I, the amplitude of the total Hall term is close to zero, as in figure 2c. 2. in region II, the spatial form of the Hall term contributed by SWs (red) fine matches with that of the total Hall term. 3. in region III, the Hall term is still dominated by SWs, but its amplitude is decreased. The main features of figure 3d are summarized as follows: 1. in region I, the PIs still dominate the Lorentz term as in figure 2d. 2. in region II, the contributions of SWs to the Lorentz term is as much as that to the Hall term in figure 3c. 3. in region III, the contribution of PIs to the Lorentz term is almost zero.

In summary, the cross shock electric field $E_x$ in region I (i.e. at the stationary extended foot) is dominated by the Lorentz term as at profile A, and is mainly contributed by PIs. In region II (i.e. at the nonstationary solar wind ion foot), the Lorentz term and Hall term are comparable to each other, and both of them are increasing. The amplitudes of the total $E_x$ and the cross shock potential in region II have already exceed that in region III. Furthermore, Hall term in this region is dominated by SWs. In region III, the amplitude of $E_x$ is decreased a lot.

\begin{figure}%[tb]
\includegraphics[width=0.95\hsize]{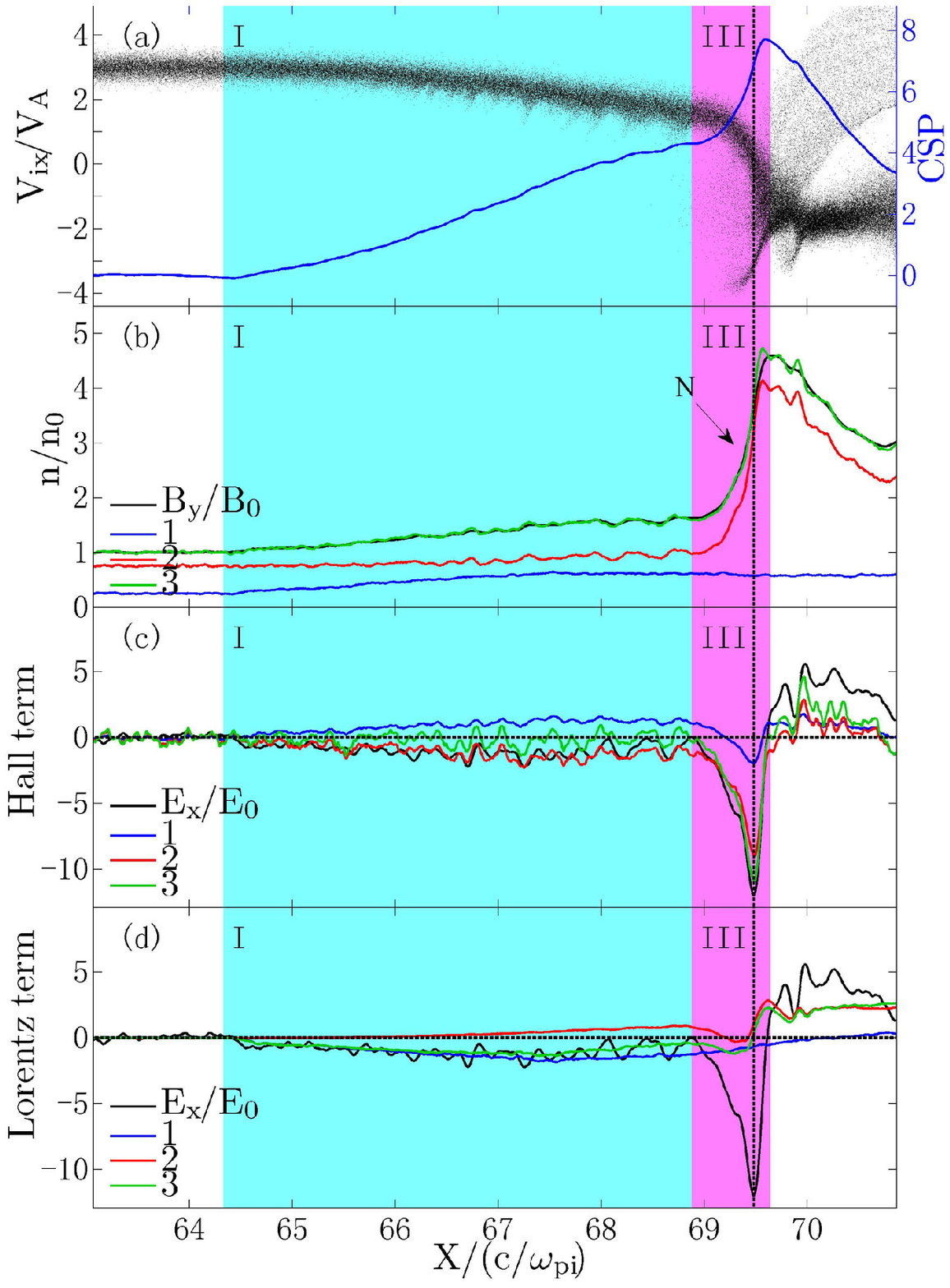}
\caption{(a) phase space plots ($V_{ix}-X$) of the solar wind ions at the shock profile A ($t=$11.3$\Omega_{ci}^{-1}$). And the cross shock potential is also shown for reference (in blue). (b) the density profiles of the pickup ions, solar wind ions and their total are in blue (line 1), red (line 2) and green (line 3), respectively. The main magnetic field $B_y$ (black curve) is also shown for reference. ``N" denotes the position of the new ramp (marked by a vertical dashed) during one reforming cycle. (c) the cross shock electric field $E_x$ is indicated by the black curve. The additional curves indicate the contributions to $E_x$ of the Hall terms on the right of Eq. (\ref{eqn:3}): pickup ion Hall term $\frac{1}{n_{e}}(\bold{J}_{PIs}\times\bold{B})_x$ (line 1), solar wind ion Hall term $\frac{1}{n_{e}}(\bold{J}_{SWs}\times\bold{B})_x$ (line 2), and their total (line 3). (d) corresponding plots for the Lorentz terms on the right of Eq. (\ref{eqn:3}): pickup ion Lorentz term $-\frac{n_{PIs}}{n_{e}}(\bold{U}_{PIs}\times\bold{B})_x$ (line 1), solar wind ion Lorentz term $-\frac{n_{SWs}}{n_{e}}(\bold{U}_{SWs}\times\bold{B})_x$ (line 2), and their total (line 3). The cyan and magenta highlight the regions: stationary foot and the nonstationary ramp, respectively.}\label{fig:4} %% no full stop at the end of caption
\end{figure}

Figure 4a-4c shows the snapshots of CSP (cross shock potential), $B_y$ and $E_x$ (in black) at shock profile C (at $t=11.3\Omega_{ci}^{-1}$) of a reforming shock, respectively. In contrast to figure 2 and 3, a striking point is that the whole shock front only can be divided into two regions because the amplitude of the new ramp (``N") have already fully overcome the old one at this time. The new ramp ``N" in region III is strong enough to reflect the incoming upstream solar wind ions, and its width is about 0.16$c/\omega_{pi}=$6.9$c/\omega_{pe}$. It located at $X=$69.48$c/\omega_{pi}$. The extended foot in region I is still stationary, and its width is about 4$c/\omega_{pi}$. Different from the results shown in figure 3, figure 4c clearly evidence that there is no region II foot ahead of the steep ramp, and the Hall term dominated by SWs perfectly matches with the total cross shock electric field $E_x$ in region III. Figure 4d shows the Lorentz term dominated by PIs apparently matches with the $E_x$ in region I.

\subsection{Effect of High Percentage Pickup Ions on the Cross Shock Electric Field (stationary case)}

\begin{figure}%[tb]
\includegraphics[width=\columnwidth]{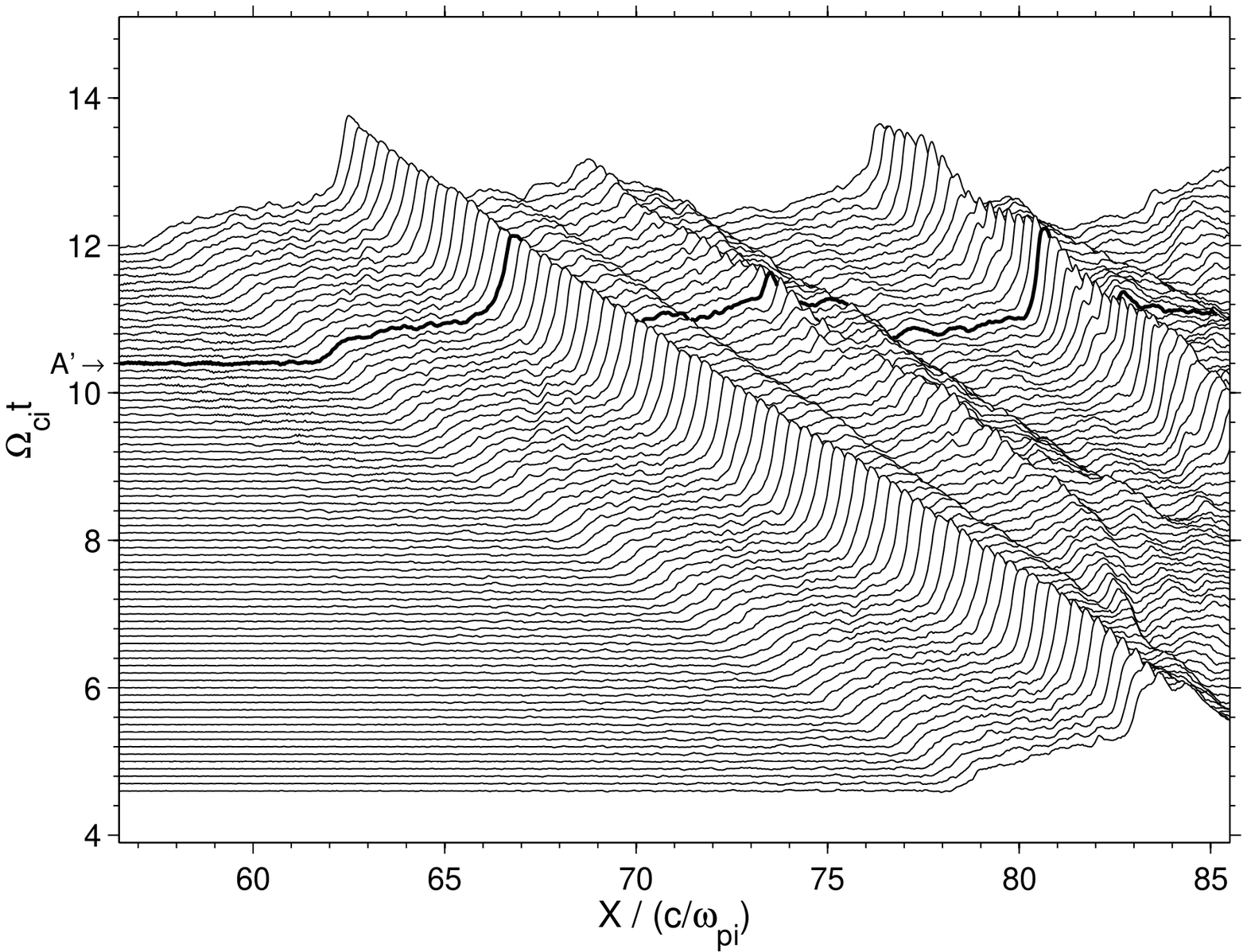}
\caption{The time evolution of $B_y$ versus $X$. Thick curve $\rm A'$ indicates the shock profile that is chosen in the stack plots at the stationary shock with 55\% pickup ions.}\label{fig:5} %% no full stop at the end of caption
\end{figure}

Secondly, let us analyze the impact of high percentage pickup ions on the cross shock electric filed at a supercritical perpendicular shock. In the current case, the percentage of the pickup ions is set to 55\%, and the other parameters of the simulation are kept unchanged. In contrast to the above results (section 3.1), a stationary shock can be demonstrated in figure 5. This stationary, supercritical shock is characterized by an overshoot, a ramp and broad foot in front of the ramp. Without loss of generality, we analyze the cross shock electric field $E_x$ at $t=$10.4$\Omega_{ci}^{-1}$ (marked by ``$\rm A'$" in figure 5) because the shock structure is not changing with time.

\begin{figure}%[tb]
\includegraphics[width=0.95\hsize]{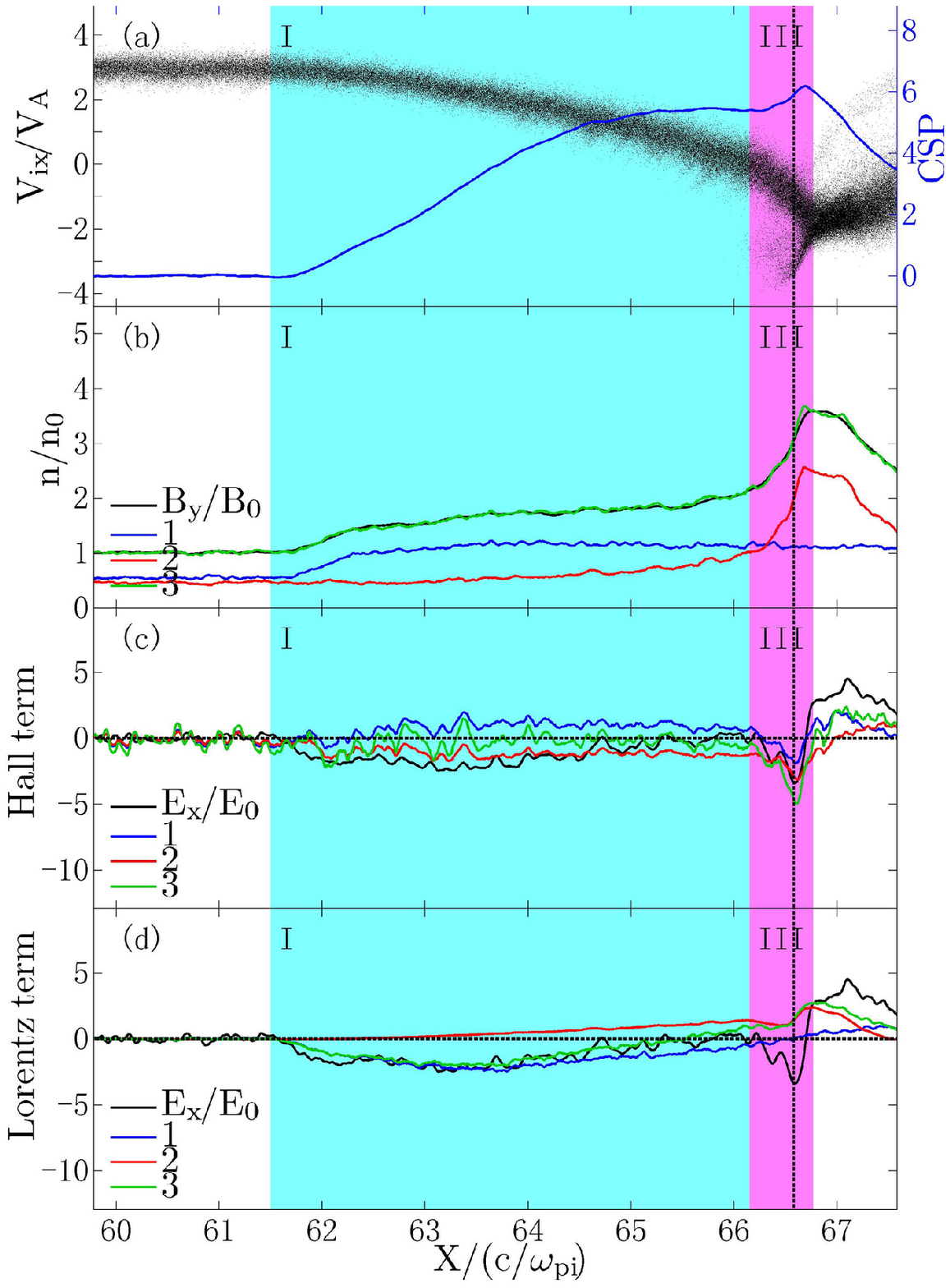}
\caption{(a) phase space plots ($V_{ix}-X$) of the solar wind ions at the shock profile $\rm A'$ ($t=$10.4$\Omega_{ci}^{-1}$). And the cross shock potential is also shown for reference (in blue). (b) the density profiles of the pickup ions, solar wind ions and their total are in blue (line 1), red (line 2) and green (line 3), respectively. The main magnetic field $B_y$ (black curve) is also shown for reference. (c) the cross shock electric field $E_x$ is indicated by the black curve. The additional curves indicate the contributions to $E_x$ of the Hall terms on the right of Eq. (\ref{eqn:3}): pickup ion Hall term $\frac{1}{n_{e}}(\bold{J}_{PIs}\times\bold{B})_x$ (line 1), solar wind ion Hall term $\frac{1}{n_{e}}(\bold{J}_{SWs}\times\bold{B})_x$ (line 2), and their total (line 3). (d) corresponding plots for the Lorentz terms on the right of Eq. (\ref{eqn:3}): pickup ion Lorentz term $-\frac{n_{PIs}}{n_{e}}(\bold{U}_{PIs}\times\bold{B})_x$ (line 1), solar wind ion Lorentz term $-\frac{n_{SWs}}{n_{e}}(\bold{U}_{SWs}\times\bold{B})_x$ (line 2), and their total (line 3). The cyan and magenta highlight the regions: stationary foot and the nonstationary ramp, respectively.}\label{fig:6} %% no full stop at the end of caption
\end{figure}

Figure 6a-6c shows the snapshots of CSP (cross shock potential), $B_y$ and $E_x$ (in black) at shock profile $\rm A'$ (at $t=10.4\Omega_{ci}^{-1}$) of a reforming shock, respectively. At this shock profile, the whole shock front can be divided into two parts: region I and region III similar to that of shock profile C mentioned in section 3.1. The shock ramp is at $X=$66.58$c/\omega_{pi}$, and its width is about 0.19$c/\omega_{pi}=$8.2$c/\omega_{pe}$. Compare with the results described in section 3.1, the amplitudes of the overshoot and cross shock electric field at shock profile $\rm A'$ (i.e. at a stationary shock) is much lower than that measured at a reforming shock, e.g. at profiles A (Figure 2) and C (Figure 4). Figure 6a shows the solar wind ions reflected at the shock ramp (marked by vertical dashed in region III) can not form a clear vortex which is used as a key evidence of shock front self-reformation [Hada et al., 2003]. Figure 6b shows the density profiles of the PIs (blue), SWs (red) and the total ions (green), as in figure 4b. In region I, the total density profile is increased, and its shape is mainly controlled by PIs. The density of SWs exceeds that of PIs in region III. Figure 6c and 6d shows the contributions of SWs and PIs to the Hall term and Lorentz term at the stationary shock (profile $\rm A'$) are analogous to that obtained at the self-reforming shock (e.g. profile C in figure 4). The only difference is that the amplitude of the cross shock electric field $E_x$ in region III is much lower, and the foot in region I is slightly higher.

In summary, high percentage PIs leads to a stationary shock (i.e. without self-reformation) even in low $\beta_{i}$ case. This point already has been pointed out by \cite{Matsukiyo2011} in which they claimed that a shock becomes stationary even with 30\% PIs. The cross shock electric field $E_x$ in region I is still dominated by the Lorentz term contributed by PIs. In region III, $E_x$ is still dominated by the the Hall term contributed by SWs. These two points are the same as that obtained at the steep and narrow shock profile (e.g. profiles A or C) in low percentage PIs case (i.e. nonstationary case). Furthermore, low electromagnetic field within the shock front can not sustain the ion vortex in phase space. Thus the region II with feature of self-reformation disappears at this stationary shock.

\section{Conclusions}

In this paper, we used 1-D PIC simulations to study the contributions of the cross shock electric field in supercritical, perpendicular shocks in the presence of pickup ions. Firstly, we separate the electric field into several terms (such as Hall term and Lorentz term), and their importance is evaluated. Secondly, we compare the contributions of pickup ions and solar wind ions to each term above. Finally, we study the impact the percentage of pickup ions on the shock front nonstationarity. The analysis has evidenced the following features:

1. In low PIs\% case, the supercritical shock is nonstationary. For the electric field ${E}_{x}$, after the new shock ramp is formed, the Hall term becomes more and more important with the increase of the new ramp. After the new ramp exceeds the old ramp, its width is very small, which is comparable with or even smaller than the ion inertial length, and the Hall term dominates the electric field ${E}_{x}$. At that time, the electric field ${E}_{x}$ is obviously larger that at other times. As pointed previously by \cite{yang2009a} the shock surfing acceleration (SSA), where the particles are  reflected mainly due to the electric field ${E}_{x}$, ion Lorentz term become negligible at that time. The characteristics mentioned above is nearly the same as that obtained at the nonstationary perpendicular shock in the absence of pickup ions \citep{yang2009b}.

2. In low PIs\% case, the extended foot caused by the reflected pickup ions also has been retrieved. $E_x$ in the extended, stationary foot region (so-called region I) is mainly contributed by the Lorentz term, which is always dominated by the pickup ions. In solar wind ion foot region (so-called region II), contributions of the pickup ions and solar wind ions are competing with each other in the Lorentz term of $E_x$. Simultaneously, the Hall term is mainly contributed by the solar wind ions. And in the ramp and overshoot (so-called region III), $E_x$ is mainly contributed by the Hall term and is always dominated by the solar wind ions.

3. In high PIs\% case, the amplitude of the extended foot is increased as expected. The striking point is that region II can not be found any more. The whole shock front becomes stationary. Because the amplitude of the overshoot and cross shock electric field is decreased and thus they are not strong enough to form the solar wind ion vortex in phase space at the ramp.

4. In high PIs\% case, there are two remainder regions (I and III) within the shock front. $E_x$ in region I is dominated by the Lorentz term contributed by the pickup ions and $E_x$ in region III is dominated by the Hall term contributed by the solar wind ions, respectively.

Recently, although the real ion-to-electron mass ratio has been employed. A few other plasma parameters are demonstrated to have great influence on the structure of the supercritical, quasi-perpendicular shock. If a high $\omega_{pe}/\Omega_{ce}$ ratio is used, a Buneman instability between reflected ions and the inflow electrons plays an important role in the foot electric field in electron inertial scale \citep{Shimada2000}, and on this scale the electron pressure term may become as important as the Lorentz term and Hall term analyzed in this paper. Simultaneously, two-dimensional PIC simulations found that a supercritical quasi-perpendicular shock may be emitted waves, such as oblique whistler waves and lower-hybrid waves within the shock front \citep{lembege2009}, which may also change the structures of the shock. In addition, Alfv\'en waves \citep{Lu2009} or upstream turbulence generated by summing many individual circularly polarized Alfv\'en waves \citep{Giacalone2010,Guo2010} are always observed in the solar wind. Clear ripples can be identified at the front of the shock after the interaction of upstream turbulence structures with a shock. How they will affect the cross shock electric field/potential and the mechanisms of ion acceleration at perpendicular shocks is our future topic.

%% Acknowledgements
%
\acknowledgments
This work was supported by the National Science Foundation of China (40974083, 40974103, 41031064, 40890164, and 40931053), Public Science and Technology Research Funds Projects of Ocean (201005017), Chinese Academy of Sciences KJCX2-YW-N28, and the Youth Fund of Polar Research Institute of China (CX20120101). Thanks are also given for the financial support of the Chinese Arctic and Antarctic Administration (10/11YR09).

%% References
%% Please cite all reference entries in the article text using \cite or
%% equivalent command.

%%%  Using BibTeX  (Name-Year style)
%
% \bibliographystyle{spr-mp-nameyear-cnd}  %% BibTeX style
% \bibliography{<bib data>}                %% BibTeX data

%% Non-BibTeX  (Name-Year style)
%

\end{document}